# Generating and Detecting High Frequency Liquid-Based Sound Resonances with Nanoplasmonics

Yanhong Wang† and Reuven Gordon*,‡

†*Science and Technology on Electronic Test and Measurement Laboratory, North University of China, No.3 Xueyuan Rd, Taiyuan, Shanxi, China, 030051*

‡*Department of Electrical and Computer Engineering, University of Victoria, Victoria, BC, Canada, V8P5C2*

E-mail: rgordon@uvic.ca

Phone: +1 250 472 5179 . Fax: +1 250 721 6052

**Abstract**

We use metal nanostructures (nanoplasmonics) excited with dual frequency lasers to generate and detect high frequency (> 10 GHz) sound wave resonances in water. The difference frequency between the two lasers causes beating in the intensity, which results in a drop in the transmission through the nanostructure when an acoustic resonance is excited. By observing the resonance frequency shifts with changing nanostructure size, the transition from slow to fast sound in water is inferred, which has been measured by inelastic scattering methods in the past. The observed behavior shows remarkable similarities to a simple Debye model (without fitting parameters). The ability to directly excite high-frequency sound waves in water may unlock the secret of how the nanofluidic environment, that is typically considered to be extremely viscous, can efficiently support the energetic dynamics of life via protein vibrations at the nanometer scale.

Water is one of the most interesting liquids to study in the extremely high frequency regime, not only because it possesses remarkable physical dynamics, but also because it supports life, which is manifest in the dynamics of proteins in that frequency regime.[1] (Note, extremely high frequency is the radio-frequency band term). The dynamics of water changes from being liquid-like to solid-like for vibrations above 10 GHz. While often overlooked, this transition is fundamental importance to biology since most globular proteins have their functional normal modes in this regime.[2]

The transition of physical properties in water have been studied by several methods. Microwave studies show that the real electric permittivity drops by a factor of 45[3] in this range. Inelastic scattering experiments (particularly neutron and x-ray) have suggested that the speed of sound doubles[4,5] and the properties of water are similar to ice.[6] The sound velocity transition has been explored with computational methods.[7] Perhaps not coincidentally with these profound changes is that the vibrational modes of proteins are found in this regime. The oscillations in solution have been probed with the optical Kerr effect[8] and optical tweezers.[9] Systematic studies have probed how solutions damp nanoparticles vibrations;[10] however, the dynamics of the liquid can be evasive.[11] Inelastic scattering has also been applied to proteins to study their dynamics in this regime,[12] but there is a lack of methods that can access the dynamics in solution.[1]

Here we use a double-nanohole (DNH) immersed in water as a means of exciting high frequency acoustic waves. The DNH is used also to detect resonances via changes in the optical transmission. We note variations in the acoustic resonances with geometry. Unlike our previous works on DNH structures, no trapping is used in this experiment. Rather, the intensity beating of two non-degenerate lasers is used to excite sound waves with the source at the cusps between the DNHs where the field intensity is the greatest. This leads to a small variation in the transmission through the aperture when an acoustic resonance is encountered. Figure 1(a) shows the experimental setup. The lasers are an external cavity tunable laser and a distributed feedback tunable laser, both operating around 850 nm, with power at the aperture of approximately 6 mW (both lasers have the same intensity).

Two possible mechanisms are considered to allow for the sound generation in this DNH geometry: heating and electrostriction. While electrostriction is certainly present in water, it is typically probed with extremely high pulse intensities.[13] Since the present experiment uses much lower peak intensities, it is believed that thermal effects dominate here. The temperature differences found for CW measurements of a similar geometry and power was approximately one degree Kelvin per mW.[14,15] This creates a refractive index variation in the liquid and modifies the transmission. Furthermore, the small cusp separation (of the order of 10 nm) means that the thermal diffusion is fast when compared with the timescales of this experiment (considering a thermal diffusitivity of approximately $10^{-7}$ m$^2$/s in water). Fast thermal relaxation has been considered in several other works at this time scale.[16–18] The DNH structures were fabricated by focussed-ion beam milling; we have also observed resonances in DNH structures fabricated by colloidal methods.[19]

Figure 1(b) shows repeated measurements of the transmission through the aperture when tuning the frequency difference (detuning) between the lasers from negative to positive frequencies. The acoustic resonances occur when the light transmitted through the aperture is at a minimum value. When on resonance, heating is greatest and the refractive index in the gap decreases on average resulting in less transmission. This is most clear for low frequency detuning (around 0 GHz). In this range, heating occurs when the lasers are in phase and the intensity is highest. This is also when the aperture transmits the least light because the refractive index is lowest, due to the same heating. When the lasers are out-of-phase, there is no heating, but there is also no light due to the destructive interference. As a result, the time-averaged transmission is reduced when averaging over the cycle, from in-phase to out-of-phase.

For higher frequencies, resonances occur when the cavity resonances are excited so that the acoustic wave returns to the gap region in-phase. Considering a sound velocity of 1.5 km/s and a dimension of approximately 100 nm, we expect these resonances to occur around the 15 GHz. The exact frequency will differ from this value due to the geometry. We also note that the positive and negative beat frequency values display the same resonances, as would be expected by the symmetry of the experiment (it makes no difference if the lasers are interchanged).

In Figure 1(b) we see that the resonance for a 164 nm diameter DNH (diameter of one of the holes) occurs at 12.7 GHz and also at -12.7 GHz. The symmetry and approximate value of this resonance are as expected from the above discussion. Also, there is a reduction in the transmission for low frequency beating (around 0 GHz) due to heating when the lasers are in-phase, as discussed above.

We systematically repeated this experiment on different aperture sizes. One would expect an inverse size dependence with zero frequency for infinitely large apertures. Figure 2(a) shows this the observed resonances. It is clear that the linear fit to the inverse size does not go through the origin. This is because the observed resonances are shifted from the natural resonances, which is typical for a driven harmonic oscillator. Simple theoretical considerations show that the natural frequency $f_0$ is related to the observed resonance frequency $f_r$ by the relation:

$$f_r = \sqrt{f_0^2 - f_0 \gamma/(2\pi)}  \quad (1)$$

where $\gamma$ is the damping rate of the resonance. Assuming that the damping is the same as found in the Debye model for water, $\gamma \simeq 1/(8 \text{ ps})$.[3] It is noted that there are typically contributions intrinsic to the nanostructure, the substrate and the environment that are considered to be less significant than the internal damping of the water since its timescale is so fast.[20–22] Figure 2(a) also shows the natural frequency value calculated from this equation, which does intercept the origin, as expected. The inset to this figure shows the scanning electron microscope images of the apertures used.

To confirm the role of the liquid in these resonances, we added glycerol to the solution and noted that the resonance frequency increased. This is consistent with the increase in the sound velocity with the addition of glycerol.[23] Figure 2(b) shows a representative curve. Other concentrations were measured as well and showed the same trend; however, the detailed investigation of mixtures and other liquids is left for a later study.

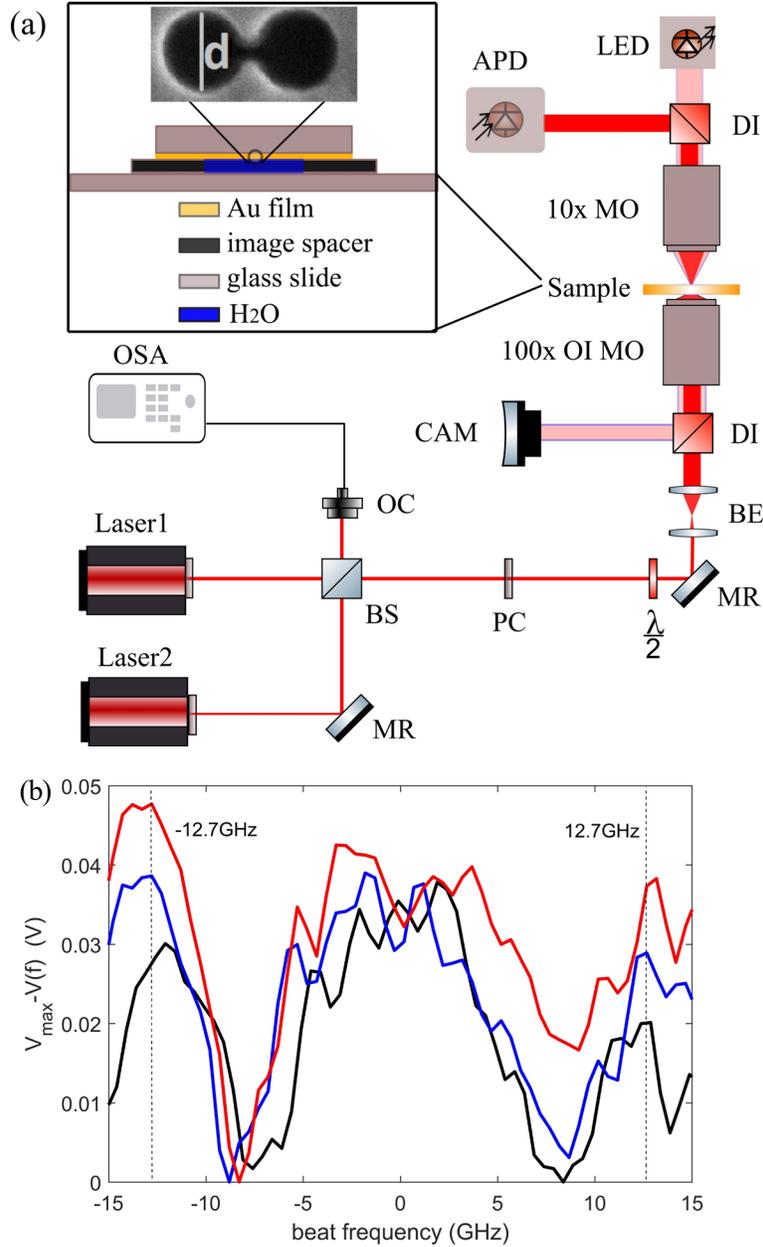

Figure 1: (a) Schematic of dual laser excitation setup of a DNH immersed in water. The intensity beating at the difference frequency between the lasers excites sound waves. Optical spectrum analyzer (OSA), optical coupler (OC), beam splitter (BS), polarizer (PC), halfwave plate ($\lambda/2$), mirror (MR), beam expander (BE), dichroic reflector (DI), microscope objective (MO), avalanche photodiode (APD), CCD camera (CAM). (b) Repeated measurements of transmission (APD voltage) for 164 nm diameter DNH as a function of beat frequency between the two lasers. The maximum value measured is subtracted to emphasize the resonances which occur at the minima of the transmission.

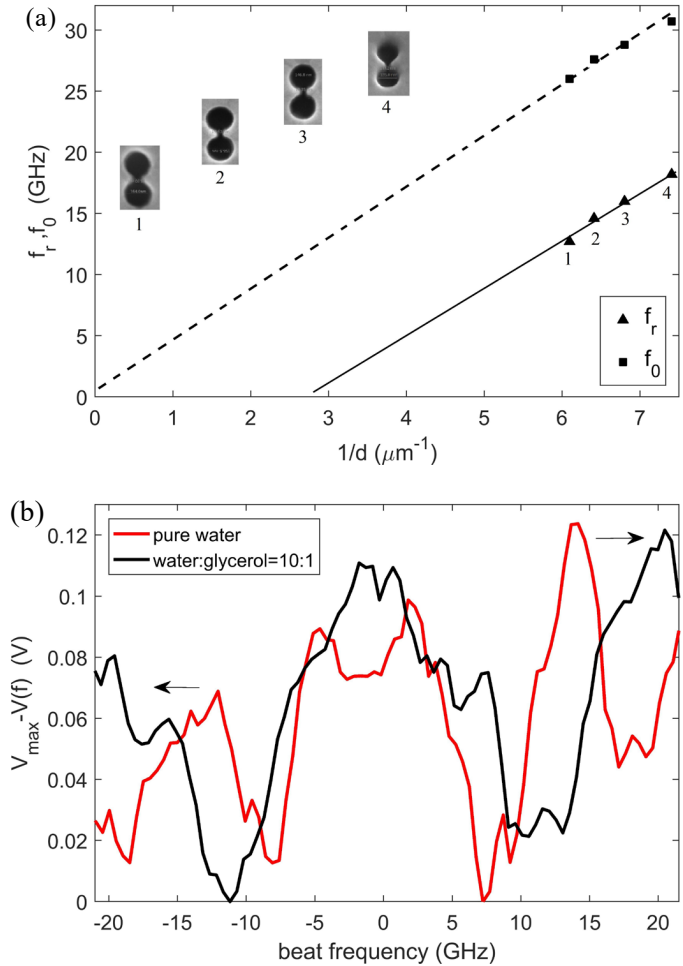

Figure 2: (a) Lowest frequency resonances measured for various aperture diameters, $f_r$. Insets shows scanning electron microscope images of measured apertures. The natural acoustic resonance $f_0$ is shifted from the measured resonance of the aperture $f_r$ since it is a driven damped oscillator. The shifted values are plotted as well as $f_0$ where the scattering time in water of 8 ps is considered to be the dominant damping mechanism. (b) Shift in resonance peaks when adding 10 % glycerol.

Figure 3(a) shows the resonances for two separate hole diameters of the DNH with resonances extending to higher frequencies. The corresponding resonances are shifted and labelled in that figure; the smaller aperture has higher frequency versions of the larger aperture. We can consider the relative changes in the frequency to estimate the change of the sound velocity of water as a function of frequency.

The resonance frequency, $f$ is expected to depend on the sound velocity in the liquid, $v(f)$, as well as the diameter of the aperture $d$ and a geometric parameter for the given resonant mode index $m$, $\alpha_m$, that depends on the shape of the aperture:

$$f = \alpha_m v(f)/d. \tag{2}$$

We can remove the geometric parameter by dividing the frequency of the larger aperture's resonances, $f$ by the frequency of the smaller aperture's resonances, denoted as $f'$:

$$\frac{f}{f'} = \frac{d'v(f)}{dv(f')}. \tag{3}$$

We can see that if the velocity of sound in the liquid is a constant, this ratio should be independent of the frequency, yet Figure 3(b) shows a dip in the ratio in the intermediate frequency regime.

We attempt to use the Debye model that has been applied to model the frequency dependence of the permittivity of water in this regime.[3] The model has a single parameter for the frequency dependence, the scattering time $\tau$, which is temperature dependent. Again, we take the scattering time to be the same as from microwave measurements of 8 ps.[3] We consider that the Debye model captures the transition from the low frequency response (speed

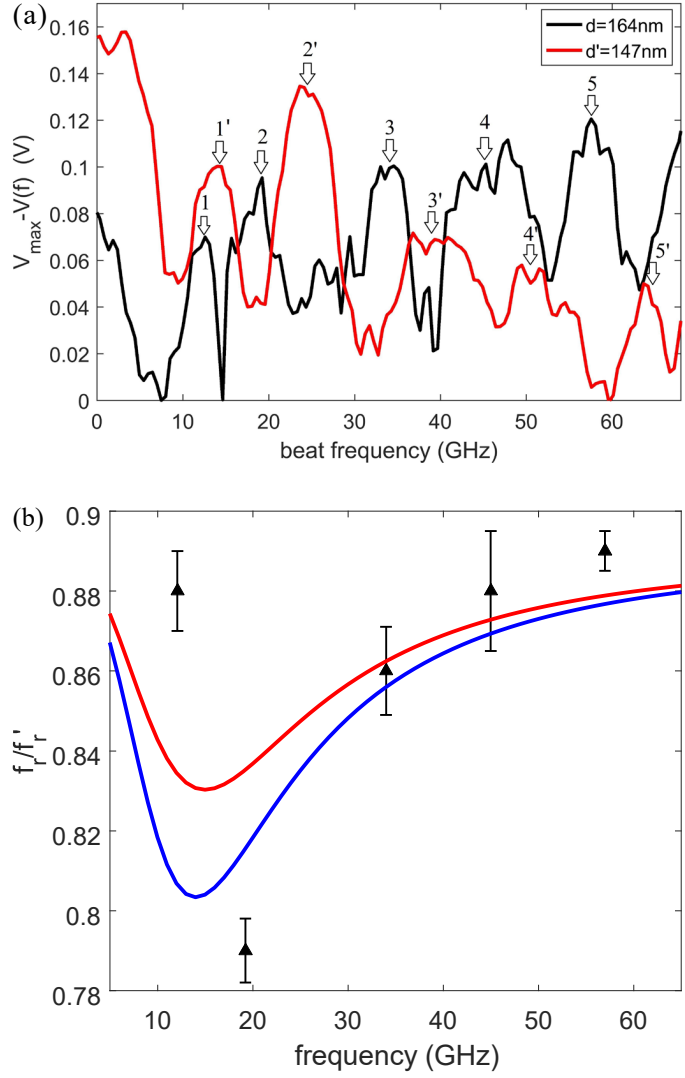

Figure 3: (a) Resonances of two separate apertures up to higher frequencies. (b) Ratio of the resonance frequencies shown in part (a) in comparison with simple Debye model (all parameters taken from other works).

of sound) to the high frequency response in a manner analogous to the real part of the permittivity:

$$v(f) = v_\infty + \frac{v_0 - v_\infty}{1 + 4\pi^2 f^2 \tau^2} \tag{4}$$

where $v_\infty$ is the high frequency limit of the sound velocity and $v_0$ is the low frequency limit.

The values for these two limits can be taken from past experiments to be 3200 m/s and 1500 m/s.[4] The results of this simple model are depicted in Figure 3(b) in red and show similarity in the shape of the curve and the minimum frequency. For curiosity, we also considered the velocity of sound in ice of 3900 m/s,[24] which is shown with the blue curve in Figure 3(b). It appears that using the sound velocity in ice provides better quantitative agreement with the experiments, but this is a discrepancy from reports using the inelastic scattering method.[6] There are clearly quantitative differences that require further investigation and more sophisticated modeling (perhaps including molecular dynamics).

In summary, here we directly measure acoustic resonances of water in the frequency regime where water transitions from being a liquid to behaving more like a solid. This measurement is enabled by nanoplasmonic excitation of the gap between the cusps in a DNH aperture in a gold film. The ability to directly excite and measure acoustic waves in water and other liquids in this regime has remarkable implications for studying the physics of liquids and their interactions with nanoparticles. In particular, it is envisioned that future work may unlock the secret of how the nanofluidic environment that is typically considered to be extremely viscous can efficiently support the energetic dynamics of life.[25,26]

## Acknowledgement

The authors thank Cuifeng Ying for providing samples when our local FIB was non-operational. The authors thank Jingzhi Wu for valuable discussions. They also thank Amirhossein Alizadehkhaledi for assistance with aligning the setup and Shahram Moradi for discussions over the resonance frequency shifts. The authors acknowledge support from the NSERC Discovery Grants program and the Chinese Scholarship Council grant 201708140170.

**1983**, *80*, 3696–3700.

(3) Kaatze, U. *Journal of Chemical and Engineering Data* **1989**, *34*, 371–374.

(4) Sette, F.; Ruocco, G.; Krisch, M.; Masciovecchio, C.; Verbeni, R.; Bergmann, U. *Physical Review Letters* **1996**, *77*, 83.

(5) Santucci, S.; Fioretto, D.; Comez, L.; Gessini, A.; Masciovecchio, C. *Physical Review Letters* **2006**, *97*, 225701.

(6) Ruocco, G.; Sette, F.; Bergmann, U.; Krisch, M.; Masciovecchlo, C.; Mazzacurati, V.; Signorelli, G.; Verbeni, R. *Nature* **1996**, *379*, 521.

(7) Rahman, A.; Stillinger, F. H. *Physical Review A* **1974**, *10*, 368.

(8) Turton, D. A.; Senn, H. M.; Harwood, T.; Lapthorn, A. J.; Ellis, E. M.; Wynne, K. *Nature Communications* **2014**, *5*, 3999.

(9) Wheaton, S.; Gelfand, R. M.; Gordon, R. *Nature Photonics* **2015**, *9*, 68.

(10) Pelton, M.; Sader, J. E.; Burgin, J.; Liu, M.; Guyot-Sionnest, P.; Gosztola, D. *Nature Nanotechnology* **2009**, *4*, 492.

(11) Chakraborty, D.; Hartland, G. V.; Pelton, M.; Sader, J. E. *The Journal of Physical Chemistry C* **2017**, *122*, 13347–13353.

(12) Doster, W.; Cusack, S.; Petry, W. *Nature* **1989**, *337*, 754.
(13) Palese, S.; Schilling, L.; Miller, R. D.; Staver, P. R.; Lotshaw, W. T. *The Journal of Physical Chemistry* **1994**, *98*, 6308–6316.

(14) Verschueren, D. V.; Pud, S.; Shi, X.; De Angelis, L.; Kuipers, L.; Dekker, C. *ACS Nano* **2018**,
10